\documentclass[aps,prb,twocolumn,showpacs,superscriptaddress]{revtex4-1}\begin{tiny}\end{tiny}
\usepackage{longtable}
\usepackage{amsfonts}
\usepackage{graphicx}
\usepackage{setspace}
\begin{document}
\title{Anharmonic Phonon Quasiparticle theory of Zero-point and Thermal Shifts in Insulators:
Heat Capacity, Bulk Modulus, and Thermal Expansion}
\author{Philip B. Allen}
\email{philip.allen@stonybrook.edu}
\affiliation{Physics and Astronomy Department, Stony Brook University, Stony Brook, NY 11794-3800, USA}
\date{\today}
\begin{abstract}
The quasi-harmonic (QH) approximation uses harmonic vibrational frequencies $\omega_{Q,H}(V)$ computed
at volumes $V$ near $V_0$ where the Born-Oppenheimer (BO) energy $E_{\rm el}(V)$ is minimum.
When this is used in the harmonic free energy, QH approximation gives a good zeroth order theory of thermal expansion 
and first-order theory of bulk modulus, where $n^{\rm th}$-order means smaller than the leading term by $\epsilon^n$, where
$\epsilon=\hbar\omega_{\rm vib}/E_{\rm el}$ or $k_B T/E_{\rm el}$,  and $E_{\rm el}$ is an electronic energy scale,
typically 2 to 10 eV.  Experiment often shows evidence for next order corrections.
When such corrections are needed, anharmonic interactions must be included.  The
most accessible measure of anharmonicity is the quasiparticle (QP) energy $\omega_Q(V,T)$ seen experimentally
by vibrational spectroscopy.  However, this cannot just be inserted into the harmonic free energy $F_H$.  In this paper,
a free energy is found which corrects the double-counting of anharmonic interactions that is made when $F$
is approximated by $F_H(\omega_Q(V,T))$.
The term ``QP thermodynamics'' is used for this way of treating anharmonicity.  It enables $(n+1)$-order corrections
if QH theory is accurate to order $n$.  This procedure is used to give corrections to specific heat and volume thermal expansion. 
The QH formulas for isothermal ($B_T$) and adiabatic ($B_S$) bulk moduli are clarified, and the route to higher-order corrections
is indicated.
\end{abstract}

%
%
\maketitle

\section{Introduction} \label{sec:intro}

In non-magnetic, insulating materials, thermodynamic behavior is controlled by vibrational excitations, which are
often close to harmonic.
There is a unique and correct version of harmonic theory, based on Taylor-expanding the Born-Oppenheimer (BO)
energy $E_{\rm el}(\{\vec{R}_{\ell}\})$ around the atomic coordinates $\{\vec{R}_{\ell,0}\}$ of a crystal with volume $V$.
The BO energy is a ``ground state'' property, and the main target of density functional theory (DFT).
This gives normal mode eigenvectors and frequencies $\omega_{Q,H}(V)$, where the index $Q= \{\vec{Q},j\}$ labels 
the states; $\vec{Q}$ runs over the $N$
wave-vectors of the Brillouin zone, and $j$ runs over the $3n$ branches.  
These states can be called ``non-interacting quasiparticles.'' However, in this paper, the term quasiparticle (QP) will
be reserved for the vibrational resonances seen experimentally.  These differ from harmonic eigenfrequencies
because higher order (``anharmonic'') terms in the Taylor expansion are not negligible.

In this paper, the term ``harmonic'' will refer to the unique correct harmonic limit, further specialized
to the case when the volume is chosen to be $V_0$, where the BO energy is minimum.  It is useful also to know the
harmonic normal modes (and their frequencies $\omega_{Q,H}(V)$) at other volumes; this is the ``quasi-harmonic'' (QH) theory.  
The harmonic approximation is a good starting point, successfully implemented by ``{\it ab initio}'' DFT
calculations \cite{Baroni},  and useful, often to good approximation, for things like specific heat, $C_P(T)$.
Vibrational spectroscopy \cite{Inel} of reasonably pure crystals most often sees reasonably sharp Lorentzian resonances.
They can be assigned a wave-vector $\vec{Q}$, and are expected to show a one-to-one correspondence with the harmonic
normal modes.  They are the QP's of this paper.
The central frequency $\omega_Q$ (the QP frequency, or energy) is $T$-dependent.
There is  good evidence from theory-experiment comparisons \cite{Nelin}
 that the QH energy $\omega_Q(V)$, evaluated at the
correct thermally-expanded volume $V(T)$ at higher $T$, does not reproduce well the measured thermal
shifts $\Delta\omega_Q$ of QP energies $\omega_Q(V,T)$.  There is an anharmonic thermal shift, additional to and
different from, the pure volume-related shift of QH theory, and it is often significant at higher $T$.

Terminology is not universally agreed upon.  Cowley \cite{oldCowley}, in his seminal paper, 
derived the modern Matsubara perturbation theory for anharmonic effects.  He occasionally
uses the word ``quasi-harmonic'' to denote what is here called ``quasiparticle''.  
In recent literature, QH most often denotes use of $\omega_{Q,H}(V)$ from $T=0$ DFT.
Occasionally papers about anharmonic theory still use ``QH'' and ``QP'' interchangeably when referring to
approximate normal modes, $\omega_Q(V,T)$, which are here called QP.

The QP relaxation rate $1/\tau_Q$ is the 
full width at half maximum of the spectroscopic Lorentzian line.  In pure crystals it lies
outside harmonic theory, and is also $T$-dependent.  
This paper is about extracting additional thermodynamic
information from the temperature dependence of $\omega_Q$, ignoring $1/\tau_Q$.
This suffices for most low-order thermal corrections.
I will call this ``quasiparticle thermodynamics''.  It differs from ``quasi-harmonic thermodynamics''.
Deviations from harmonic vibrations are responsible for thermal shifts of many properties.  The
ones of prime concern in this paper are $C_P$ and $C_V$ (constant pressure and constant volume), 
bulk modulus $B_T$ and $B_S$ (isothermal and adiabatic), and volume expansion $V(T)$ and 
$\alpha=(1/V)dV/dT$ (constant pressure.)  Good general discussions are in older literature.
\cite{Born,Leibfried,oldCowley,WallaceOld,Bilz}

There are two main ideas in QP theory: (1) low-lying excitations correspond 1-to-1 with a
non-interacting single-particle picture; they have QP energies
$\hbar\omega_Q(V,T)$ and mode occupancies $\langle \hat{n}_Q \rangle$; 
and (2) low energy dynamics can be described as the dynamics, in space and time, of the mode occupancy.
QP theory can fail in at least two ways.  (a) The resonance can be very non-Lorentzian so that $\omega_Q$ is poorly
defined.  (b) Correlated occupancy $\langle \hat{n}_Q \hat{n}_{Q^\prime} \rangle - \langle \hat{n}_Q \rangle 
\langle \hat{n}_{Q^\prime} \rangle$ may become important.  In this paper, $n_Q$ 
denotes the equilibrium (Bose-Einstein) mean occupancy $[\exp(\hbar\omega_Q(V,T)/k_B T) -1]^{-1}$,
$n_{Q,H}$ its harmonic or QH version, and
$\langle \hat{n}_Q \rangle$ denotes the actual occupancy in a general ensemble, not necessarily equilibrium.  
Entropy plays a special role\cite{Landau0}, since it is just $1/N$ times 
the logarithm of the number of ways of distributing $\langle n \rangle N\hbar\omega$ of excitation
energy into $N$ oscillators of frequency $\omega$,
\begin{equation}
\hat{S}=k_B \sum_Q [(\langle \hat{n}_Q \rangle +1)\ln(\langle \hat{n}_Q \rangle+1)-\langle \hat{n}_Q \rangle
\ln \langle \hat{n}_Q \rangle].
\label{eq:Shat}
\end{equation}
The equilibrium occupancy $n_Q$ is the one which maximizes Eq.(\ref{eq:Shat}) at fixed energy, and
the thermodynamic entropy $S(T)$ is given by Eq.(\ref{eq:Shat}) with $\langle \hat{n}_Q \rangle \rightarrow n_Q$.
When harmonic frequencies $\omega_{Q,H}$ are used in $n_Q$, the result is the harmonic entropy $S_H$.  When
the $T$-dependent QP energies are used in $n_Q$, and inserted in $S(T)$, an accurate improvement of the
thermodynamics is achieved.  This will be denoted $S_{\rm QP}$.
The same replacement does not work for the harmonic free energy.  If $\omega_Q(T)$ is inserted into $F_H$,
the resulting $F$ does not obey $-dF/dT = S_{\rm QP}$.  A ``correct'' QP free energy that does agree with $S_{\rm QP}$
is found as follows.  The thermodynamic energy $U(T)=F+TS$ in harmonic theory is 
\begin{equation}
U_H = \sum_Q \hbar\omega_{Q,H} (n_{Q,H}+1/2).
\label{eq:UH}
\end{equation}
When normal mode frequencies acquire an anharmonic correction, $\omega_{Q,H} \rightarrow \omega_Q(T)$,
the energy acquires a correction,\cite{Abanov}
\begin{eqnarray}
U_{\rm QP} &=& \sum_Q \hbar\omega_Q(T) (n_Q+1/2) \nonumber \\
&-&(1/2)\sum_Q \hbar[\omega_Q(T)-\omega_{Q,H}](n_Q+1/2)
\label{eq:UQP}
\end{eqnarray}
This corrects for double-counting of the interaction, but only in leading anharmonic approximation where \cite{WallaceOld}
\begin{equation}
\omega_Q(T)-\omega_{Q,H} \equiv \Delta_Q^{(2)}
= \frac{1}{N}\sum_{Q^\prime} \left(\frac{\partial\omega_Q}{\partial n_{Q^\prime}}\right)_0 
\left(n_{Q^\prime}+\frac{1}{2}\right),
\label{eq:anhsh}
\end{equation}
where the superscript $\Delta^{(2)}$ indicates the lowest order correction (second-order perturbation theory) which
comes from third and fourth-order anharmonicity.
Further details are in the Appendix.  Here $\partial\omega_Q/\partial n_{Q^\prime}$
is a $T$-independent anharmonic interaction function.  In higher-order perturbation theory there are presumably
additional shifts of the type
\begin{equation}
\Delta_Q^{(3)}
= \frac{1}{N^2}\sum_{Q^\prime,Q^{\prime\prime}} \left(\frac{\partial^2\omega_Q}{\partial n_{Q^\prime}\partial n_{Q^{\prime\prime}}}
\right)_0 \left(n_{Q^\prime}+\frac{1}{2}\right)  \left(n_{Q^{\prime\prime}}+\frac{1}{2}\right)
\label{eq:anhsh3}
\end{equation}
involving anharmonic interactions up to sixth-order in displacement $u_Q$.  We assume these can be omitted, and
this approximation enables the correction in Eq.(\ref{eq:UQP}) to be sufficient.  Then an accurate
and consistent thermodynamic free energy is $F_{\rm QP}=U_{\rm QP}-TS_{\rm QP}$. 
Notice that the anharmonic shift, Eq.(\ref{eq:anhsh}), does not vanish at $T=0$, but has a zero-point component
where $n_{Q^\prime}+1/2\rightarrow 1/2$.  The quasiparticle frequencies are shifted from the harmonic frequencies
even at  $T=0$.

QH thermodynamics
is a limiting case.  It uses only volume-dependence of $\omega_{Q,H}(V)$.  The correction factor in $U_{\rm QP}$,
Eq.(\ref{eq:UQP}), vanishes, and the quasi-harmonic free energy 
is just the harmonic free energy with a volume-dependent harmonic frequency. 
It improves pure harmonic theory and gives correct lowest-order thermal
shifts for properties such as the bulk modulus which are volume derivatives of $F$.  The reason it works to lowest order is
because $d\omega_Q/dV$ differs from $d\omega_{Q,H}/dV$ only in next order.  QH theory is computationally accessible, but QP theory much less so.   QP theory suffers from the difficulty
that the anharmonic shift is not usually measured except at a few temperatures.  It can be numerically computed
using DFT for the anharmonic forces.  It is not yet computed routinely, but this is changing 
\cite{Liu1,Liu2,Hellman1,Hellman2,Hellman3,Hellman4}.

Exact theory associates vibrational
resonances with poles of a phonon Green's function, a correlation function describing dynamics on the BO energy surface.
Exact thermodynamics should be computed from a corresponding
theory for the free energy.  This can be computed perturbatively 
\cite{oldCowley,Doniach}.  At high $T$, classical molecular dynamics (MD) describes dynamics non-perturbatively, if the
BO forces are known.  This is called {\it ab initio} MD, or AIMD.  
Thermodynamics generally then requires a tricky ``thermodynamic integration'' 
\cite{Hellman3,Souvatzis,Thomas,Wentzcovitch,Errea}. 
To do a correct non-perturbative computation at lower $T$ requires quantum corrections, as in path-integral MD.\cite{Herrero}

Zero-point and related thermal nuclear motions cause shifts 
and isotope-dependences in many measured physical properties 
\cite{Born,Leibfried,oldCowley,WallaceOld,Grimvall,DAnderson,OAnderson,Wallace,Cardona,Allen}.
Explicit formulas are given here for the first-order shifts of $C_P$, $B_T$, $B_S$, and $\alpha$.
If no specification (adiabatic 
{\it versus} isothermal) is made, isothermal is implied.  The adiabatic shift can be
found by thermodynamic rules, as shown in Sec. \ref{sec:BM}. 
Many of these results can be found in some form in the literature.  There is a lot of correct, 
\cite{Born,Leibfried,oldCowley,WallaceOld} plus much partially correct, 
as well as incorrect or confusing literature on this subject.  There are semi-empirical formulas that have
a long history of enabling useful fitting, even though the formulas do not seem to be justifiable in 
detail. \cite{OLAnderson,Varshni}  
The aim of this paper is a simplified, possibly less
confusing, derivation of correct formulas.

The paper is organized as follows.  
In Sec. \ref{sec:graphs}, examples of thermal shifts from experiments are given.
In Sec. \ref{sec:NC}, extra complexities of non-cubic crystals, and crystals with internal coordinates, are discussed.
In Sec. \ref{sec:QHQP}, the QH approximation and the QP approximation
are discussed.  Specific heat formulas are presented, showing that QP theory
provides a thermal correction inaccessible in QH approximation.
In Sec. \ref{sec:vol}, two orders of thermal correction to the volume are discussed.  This gives Gr\"uneisen
theory of thermal expansion $\alpha_0$ plus a first order correction. 
In Sec. \ref{sec:BM}, the leading correction to the bulk modulus is derived (from QH theory).
The Appendix reviews the microscopic theory of Eqs.(\ref{eq:Shat}-\ref{eq:anhsh}).

\section{Experimental examples} 
\label{sec:graphs}

\begin{figure}[h]
\includegraphics[width=0.4\textwidth]{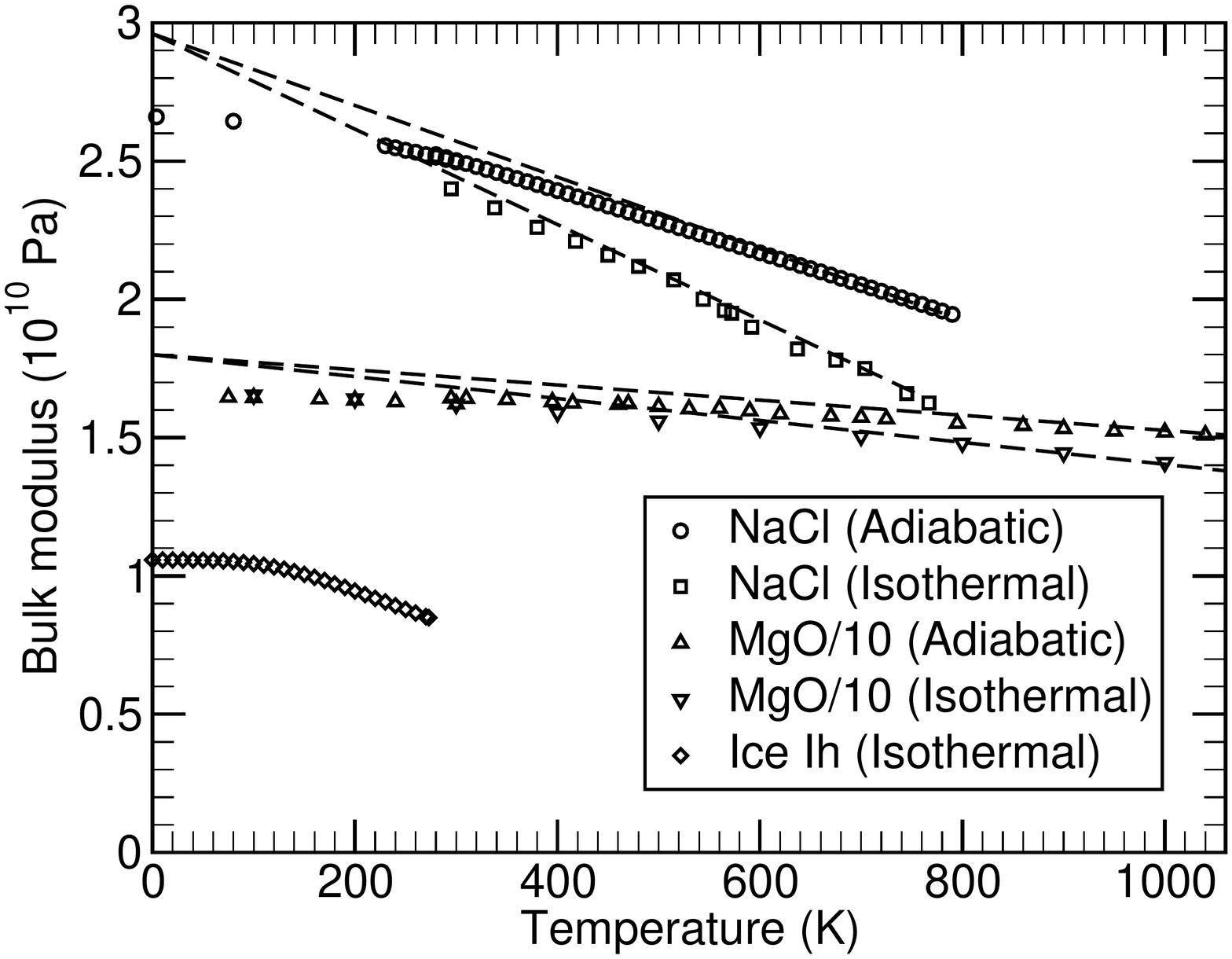}
\caption{Experimental bulk modulus B versus temperature.  Isothermal data for ice Ih are from ref. \onlinecite{Feistel}.
For NaCl, the low $T$ adiabatic data are from ref. \onlinecite{Lewis}, and high $T$ data are from ref. \onlinecite{Slagle}.
The isothermal data for NaCl are from ref. \onlinecite{Yamamoto}.
For MgO, adiabatic low $T$ data are from ref. \onlinecite{And-And} and high $T$ data from ref. \onlinecite{Soga}.
Isothermal data for MgO are from ref. \onlinecite{Isaak}.
The MgO data are larger by 10 than the others, and have a weaker thermal shift.
Many metals (Al, Cu, Ni, {\it etc.}, ref. \onlinecite{Wang}) have bulk moduli similar in magnitude to MgO, with
large thermal shifts (similar to NaCl, for $\Delta T \sim \Theta_D$).
Covalent materials, like carbon \cite{Mounet} and silicon \cite{Nikanorov}, 
are more similar to MgO, showing weaker $T$-dependence .
The dashed lines are approximate extrapolations suggesting that the zero-temperature, frozen-lattice
values of B are larger than the measured zero-temperature values by $\sim 15\%$ (NaCl) and $\sim 9\%$ (MgO).
The extrapolation for ice Ih is not given, and would require detailed calculation as explained in the text.}
\label{fig:data}
\end{figure}

\begin{figure}[h]
\includegraphics[width=0.4\textwidth]{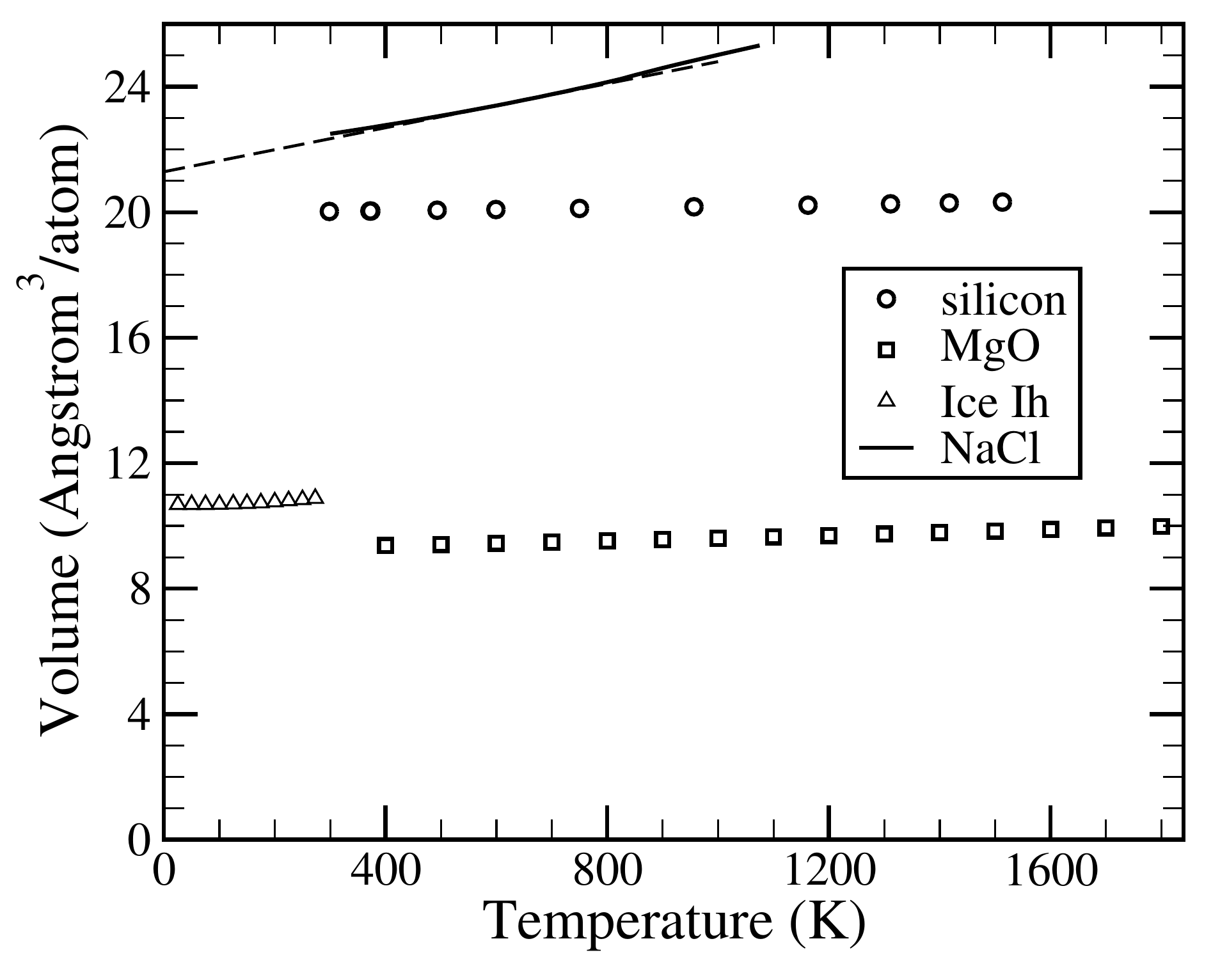}
\caption{Experimental volume $V_a$ per atom versus temperature.  
Data for silicon are from ref. \onlinecite{Okada}.
Data for MgO are from \onlinecite{Shanker}.
Data for ice Ih are from ref. \onlinecite{Feistel}.
Data for NaCl were constructed by integrating the polynomial expressions for linear thermal expansion
given in ref. \onlinecite{Enck}.  The dashed curve extrapolates the NaCl $V(T)$ using the quasi-harmonic
high $T$ slope (valid for $T>\Theta_D$) $dV_a/dT=3k_B \bar{\gamma}/B_0$, with experimental
values $\bar{\gamma}=1.57$ and $B_0=23.7$GPa, as tabulated in ref. \onlinecite{Matsui}.  This suggests
that the zero-point expansion $\Delta V(T=0)/V_0$ of NaCl is about 5\%.  Crude extrapolations for 
MgO and silicon suggest zero-point expansions of less than 2\% and 1\% respectively. 
The extrapolation for ice Ih is not given.  Detailed calculations for ice show that (as is true for silicon as well) Gr\"uneisen
parameters are negative for some modes and positive for others; the theoretical zero-point expansion
of ice Ih was computed to be $\approx 1$\% in ref \onlinecite{BPamuk}.}
\label{fig:vol}
\end{figure}

\begin{figure}[h]
\includegraphics[width=0.4\textwidth]{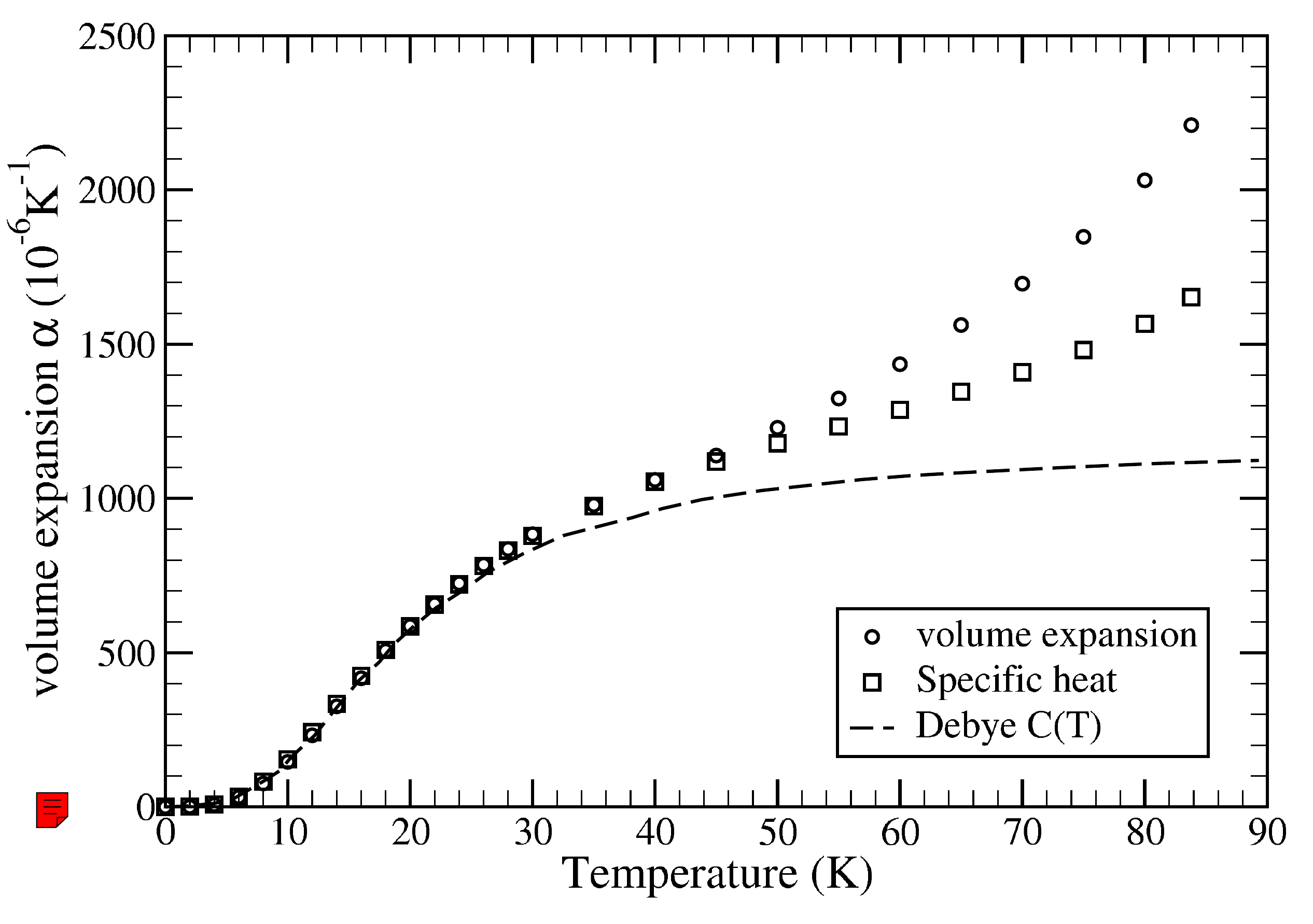}
\caption{Experimental volume expansion $\alpha(T)$ and specific heat $C_P(T)$, for crystalline argon,
at $P=$1 atmosphere, up to the melting temperature (82.3K).  The data were compiled by Bodryakov \cite{Bodryakov},
who analyzed $\approx10$ different experiments.  Shown for comparison is the Debye model with
$\Theta_D = 81.2$K fitted to $C_P$ data. Both experimental and Debye-model specific heats
were scaled, by the same factor, to lie on top of the volume expansion curve at lower $T$.}
\label{fig:alpha}
\end{figure}

Figures \ref{fig:data},\ref{fig:vol},\ref{fig:alpha} illustrate the thermal shifts under discussion.  Fig. \ref{fig:data}
shows that the bulk modulus has surprisingly large vibrational corrections
\cite{Feistel,Lewis,Slagle,Yamamoto,And-And,Soga,Isaak,Wang,Mounet,Nikanorov}.  These have serious
significance for geoscience, for example.\cite{OAnderson,DAnderson}  The leading-order bulk modulus, $B_0$, comes from 
electronic stiffness.  The product $B_0 V_a$ ($V_a=V_0 / Nn$ is the volume per atom in leading order theory)
has order of magnitude 10 eV, a characteristic electron energy.  Vibrational energies are two orders of magnitude
smaller.  I will define $\epsilon$ as the dimensionless ratio of phonon to electron energies.  This will appear
explicitly in the form $\hbar \omega/BV_a$ or $k_B T/BV_a$ in various results.  
A parameter like $\epsilon\approx 0.01$ controls the size 
of the vibrational corrections under discussion.  Fig. \ref{fig:data} shows $\approx 10\%$ shifts, indicating that 
there can be a significant prefactor multiplying $\epsilon$.

 The small parameter $\epsilon$
can be estimated as $\epsilon\approx k_B \Theta_D/B_0 V_a$,
where $\Theta_D$ is the Debye temperature.  Experimental $B$ and
$V_a$ may be used.  Rough values are $\epsilon = 0.0045$ (silicon
\cite{Nikanorov}), 0.0087 (MgO\cite{Landolt}), 0.0072 (NaCl\cite{Landolt}), and 0.042 (ice Ih\cite{Feistel,Gagnon}).
However, the parameter $\epsilon$ for ice Ih is poorly defined.  The number $\epsilon$=0.042 used the
low $T$  $\Theta_D \approx 300$K.  This measures only thermally excited (acoustic and librational)
vibrations at $T\le$ 273K.  The ``O-H
stretch'' vibrations at the opposite end of the spectrum have $\hbar\omega/k_B$ larger by 11.  These modes also
contribute to the zero-point shifts in ice.  If optic modes are used to define $\epsilon$, the value of 
this ``small parameter'' increases to 0.5.

It is perhaps worth mentioning that the representation of a physical property $P$ as an expansion in $\epsilon$  
($P=P_0 + P_1 \epsilon + P_2 \epsilon^2 + ...$) is not forced to have universally-defined coefficients.  Especially
because the expansion is truncated after the $P_1$ or possibly $P_2$ term, it is normal that the last coefficient
may (or may not, depending on the source) contain some higher effects ($P_1 = P_{10}+P_{11}\epsilon$, for example.)  
The only rule is that $P_n \epsilon^n$
should contain nothing of lower order than $\epsilon^n$.  This will be mentioned again in Secs. \ref{sec:vol} and \ref{sec:BM}. 

The volume shifts \cite{Feistel,Okada,Shanker,Enck,Matsui,BPamuk} shown in Fig. \ref{fig:vol} are smaller
in relative size.  Fig. \ref{fig:alpha} shows that $\alpha$, the temperature derivative of the volume \cite{Bodryakov}, roughly
follows a harmonic specific heat ($C_{\rm H}$) type of $T$-dependence.  This is the result of Gr\"uneisen theory \cite{Gruneisen}.
But at higher $T$, there is a very significant thermal shift of both $C_P(T)$ and $\alpha(T)$ away from the $C_{\rm H}(T)$ form.

Figs. \ref{fig:data} and \ref{fig:vol} also illustrate zero-point shifts.  Mean square thermal lattice displacements
of the $i^{\rm th}$ atom, in harmonic theory, are\cite{Born} 
\begin{equation}
\langle u_i^2 \rangle= \sum_Q\frac{\hbar}{M_i\omega_{Q,H}}| \langle i|Q\rangle|^2 \left( n_{Q,H}+\frac{1}{2} \right),  
\label{eq:u2}
\end{equation}
where $\langle i|Q \rangle$ is the component of the phonon eigenvector $|Q\rangle$ on the $i^{\rm th}$ atom, and
where $n_Q$ is the equilibrium occupation number.
The zero-temperature value $\langle u^2 \rangle \sim \hbar/2M\omega$ is
the quantum zero-point motion, which depends on nuclear mass, whereas the high-$T$ value $k_B T/M\omega^2$
is classical and depends only on the force constant $M\omega^2$, not on the nuclear mass $M$.  The low $T$
$\langle u^2 \rangle$ causes zero-point shifts of atomic volume $V(T=0)$ and bulk modulus, which differ for
different isotopes.  Therefore,
the DFT (``frozen-lattice'') value $V_0$ or $B_0$ should differ from the actual value $V(0)$ or $B(0)$.  It is interesting that
the frozen-lattice value can sometimes be deduced from experiment.\cite{Allen}  This is because the thermal factor $n+1/2$
at high $T$ becomes $x(1-1/12x^2 + \cdots)$ where $x=k_B T/\hbar\omega$.  An asymptotic linear-in-$T$ fit to
$n+1/2$ at high $T$ passes through $0$ at $T=0$.  It is of course difficult to find the ``correct'' experimental
asymptote, since thermal corrections enter to alter it.  However, the bulk modulus
simplifies the fit if both isothermal and adiabatic versions are available, because each
should extrapolate to the same $T=0$ value $B_0$, as is shown on Fig. \ref{fig:data}.  Curves of
this type are in the review of Leibfried and Ludwig. \cite{Leibfried}

We have already seen that the anharmonic shift $\Delta\omega_Q=\omega_Q-\omega_{Q,H}$
of phonon frequencies has a similar form, Eq.(\ref{eq:anhsh}).
The $T$-independent coupling parameter 
$\partial \omega_Q/\partial n_{Q^\prime}$ has contributions like $V_4$ and $|V_3|^2/\hbar\omega$;
see the Appendix, Eq.(\ref{eq:G3}).
They have order-of-magnitude $\epsilon\hbar\omega_Q$.  For example, the term 
of the type $|V_3|^2/\hbar\omega$ involves a third order anharmonic coupling coefficient $V_3$, of structure
$u^3\partial^3 E_{\rm el}/\partial u^3$, and of order $(u/a)^3 E_{\rm el}$.  The ratio $(u/a)^2$ of lattice
displacement to interatomic distance is of order $\hbar\omega/M\omega^2 a^2 \approx\epsilon$.  Putting
all these factors together, we see that $\partial\omega_Q/\partial n_{Q^\prime}\approx\epsilon\omega_Q$
(or $\Delta\omega_Q/\omega_Q\approx\epsilon$.)

The smallness of the anharmonic shift is only a crude estimate which sometimes may fail.    A failure is likely
to cause anharmonic broadening $\Gamma_Q$ of vibrations to be bigger than the spacing of
vibrational levels $\omega_Q$.  In such cases, phonon quasiparticles are poorly defined,
perturbative treatments may fail, and thermal shifts may not be described well by quasiparticle theory.

The validity of perturbative computation beyond harmonic approximation for thermodynamic properties is not a closed issue.
Wallace \cite{Wallace} summarizes evidence for failure of anharmonic perturbation theory to reproduce apparently reliable
MD.  But Boltzmann equation treatments of thermal conductivity are now very successful \cite{Omini,Ward,Li}, and are based on the third-order term in the same perturbation theory.   Computations based on DFT anharmonic forces are becoming
more common, and generally claim decent agreement with experiment.  A nice example is theory and 
experiment for aluminum by Tang {\it et al.} \cite{Tang}.
A thermal conductivity $\kappa > $ 5 W/mK is a good hint that phonon quasiparticles are mostly 
well behaved.  This crude estimate comes from $\kappa \sim Cv\ell/3$ where the specific heat is $C=3k_B/\Omega_a$,
$\Omega_a$ is the volume per atom, $v\sim \pi\omega_{\rm max}/a$ is the sound velocity, $a$ the lattice constant,
and $\ell$ the phonon mean free path.  Quasiparticle theory requires $\ell$ large compared to $a$, or $\kappa$
large compared to $\kappa_{\rm min}=Cva/3$.  If we choose $\hbar\omega_{\rm max}/k_B\sim\Theta_D$ to be 300K, and
$a$ to be 5$\AA$, then $\kappa_{\rm min} \sim 1$W/mK, and $\kappa>5\kappa_{\rm min}$ should be sufficient
to trust quasiparticle theory for most of the phonons of the material.  However, $\Theta_D$ may be significantly
bigger or smaller than 300K, and the criterion could be scaled to $\kappa > (\Theta_D/300{\rm K})\times$ 5 W/mK.

\section{Non-cubic crystals and internal coordinates}\label{sec:NC}

Pressure, volume, and temperature are not the only thermodynamic variables in crystals.
One can also have anisotropic stress $\sigma_{\alpha\beta}$ and anisotropic strains $\epsilon_{\alpha\beta}$.
Pressure and volume change are the traces of these tensors.  This paper looks only at
pressure and volume.  The generalization to tensor properties complicates notations and results, but the principles
are not changed.  Consider hexagonal structures as a simple example of non-cubic.  The separate $a(T)$ and
$c(T)$ lattice parameters are relevant thermodynamic variables.  They are also not considered in this paper,
only $V(T)=\sqrt{3}a^2 c/2$ is considered.  When $T$ changes, not only does $V$ change, but also $c/a$.
This can not be ignored, but is kept hidden in this paper.  The volume-dependent phonon frequency $\omega_{Q0}(V)$
is treated as a well-defined variable.  There is a hidden assumption that this has been computed at various volumes,
and for each volume, the correct $c/a$ ratio has been found and used in the phonon calculation.  Finally, consider
wurtzite crystal structure (hexagonal symmetry and 4 atoms per cell).  There is one ``internal coordinate'' $u(T)$,
which determines the $c$-axis offset between the cation and anion sublattices.  This also cannot be ignored.
But it is hidden by the assumption that for a particular choice of $V$, the correct $u(T)$ as well as $c/a$ have
been computed, and used to find $\omega_{Q,H}=\omega_Q(V_0)$, and $\partial \omega_{Q,H}/\partial V$, {\it etc.}
Cubic crystals can also have internal coordinates not fixed by symmetry, which need to be treated the same way.

\section{Quasiparticle thermodynamics} \label{sec:QHQP}

Even when harmonic
approximation is seriously perturbed by anharmonic effects, there may still be phonon
quasiparticles, with effective interactions not too strong.  Thermodynamics is then 
approximated by using the QP frequencies $\omega_Q(V,T)$ in the non-interacting entropy formula\cite{Landau0},
\begin{equation}
S_{\rm QP} = k_B \sum_Q \left[(n_Q + 1)\ln(n_Q +1) -n_Q \ln n_Q \right].
\label{eq:Sqp}
\end{equation}
Because of the $V$ and $T$-dependence
of the QP energy, $S_{\rm QP}$ has altered $V$ and $T$-derivatives which give
corrections in thermodynamic calculations \cite{WallaceOld,Grimvall,Barron,Hui,Cowley0}.
The corresponding free energy is
\begin{equation}
F_{\rm QP}=E_{\rm el}(V) +  F_{\rm vib,H}(V,T) + \Delta F_{AH},
\label{eq:Fqp}
\end{equation}
\begin{equation}
F_{\rm vib,H}=k_B T\sum_Q  \ln\left[2\sinh\left(\frac{\hbar\omega_Q(V,T)}{2k_B T}\right)\right]
\label{eq:Fvib}
\end{equation}
\begin{equation}
\Delta F_{AH}=-\frac{\hbar}{2}\sum_Q \Delta_Q^{(2)} (n_Q+1/2),
\label{eq:DF}
\end{equation}
where $\Delta_Q^{(2)}=\omega_Q(V,T)-\omega_{Q,H}$ was defined in Eq.(\ref{eq:anhsh}).
The part $F_{\rm vib,H}$ has the standard form of the harmonic free energy, but here the quasiparticle
frequency $\omega_Q(V,T)$ is inserted.  The double-counting correction $\Delta F_{AH}$ is nominally
smaller by $\epsilon$ than the part $F_{\rm vib,H}$.
This version of $F_{QP}$ is the same as $U_{QP}-TS_{QP}$ and Eq.(\ref{eq:UQP}) for $U_{QP}$.
The quasi-harmonic formulas $S_{QH}$ and $F_{QH}$ are the same except that 
$\omega_Q(V,T)$ is replaced by $\omega_Q(V)$, usually calculated by DFT.
In that case, the anharmonic term, Eq.(\ref{eq:DF}), vanishes.
In a metal or a magnetic material, one should
include additional terms in $F_{\rm QP}$ for thermal excitation of electrons or magnons.  
Such effects are omitted here.  
The QH procedure of using just a volume-dependent
QP energy in the harmonic free-energy formula, does give correct first-order $V$-derivatives, but fails to give 
thermal shifts which depend on $T$-derivatives.  In this sense, it can be regarded as an
incomplete, rather than an incorrect, theory, and as a partially
correct simplification of QP theory.  It correctly contains the information available from
DFT calculations of the frequency-spectra at different volumes.  Ramirez {\it et al.} \cite{Ramirez} made
a careful study of the accuracy of the QH approximation by comparison with well-converged path-integral MD
for three phases of ice.  They find generally very good agreement between QH and PIMD

As an example of QP thermodynamics, consider the specific heat, $C=T dS/dT$.  The free energy is not needed;
the correct QP entropy is Eq.(\ref{eq:Sqp}) with QP frequencies in the equilibrium occupation functions.
\begin{equation}
C_X=T\left(\frac{\partial S}{\partial T}\right)_X \approx \sum_Q\hbar\omega_Q\left(\frac{\partial n_Q}{\partial T}\right)_X.
\label{eq:Cp0}
\end{equation}
Here $X$ is pressure $P$ or volume $V$.  This gives
\begin{equation}
C_{X,{\rm QP}}=\sum_Q \hbar\omega_Q \left(\frac{dn_Q}{dT}\right)_{\rm H} \left[1-\frac{T}{\omega_Q}
\left(\frac{\partial \omega_Q}{\partial T}\right)_X \right],
\label{eq:Cp1}
\end{equation}
where the subscript ``H'' means $(dn_Q/dT)_{\rm H}=(\hbar\omega_Q/k_B T^2)n_Q(n_Q+1)$,
obtained by differentiating $n_Q$ by the explicit $T$ in the Bose function, but not by the
implicit $T$ contained in $\omega_Q(V,T)$.
The first term of Eq.(\ref{eq:Cp1}) is a harmonic specific heat $C_{\rm H}$, but 
not the purely harmonic $C_0$, because the frequencies
$\omega_Q$ appearing in the formula are the renormalized $T$-dependent quasiparticle frequencies.
The difference between $C_{\rm H}$ and $C_0$ is a gentle $T$-dependent stretching of the harmonic 
$C_0(T)$ curve along the $T$ axis.  This does not affect the high $T$ classical limit, $3Nnk_B$.
A serious high-$T$ deviation from harmonic theory (see the measurement for Ar, Fig. \ref{fig:alpha}) must
be caused by the second term of Eq.(\ref{eq:Cp1}).

In QH theory, $(\partial\omega_Q/\partial T)_V=0$, so $C_{V,{\rm QH}}=C_{\rm H}$.
Also in QH theory, $(\partial\omega_Q/\partial T)_P=(\partial\omega_Q/dV)_T (\partial V/\partial T)_P$,
so there is a significant QH correction to $C_{P}$.
QH theory gives the correct difference, $C_P  - C_V$, but it misses the anharmonic
correction which appears in the correct QP 	theory for both $C_V$ and $C_P$.
Computational evidence \cite{Hui} shows 
that QH theory shifts $C_P(T)$ away from the harmonic value $C_0$, 
but that experiment exhibits different shifts.\cite{Karki}

\section{Volume expansion } 
\label{sec:vol}

The aim is to get corrections to one higher order than the standard Gr\"uneisen theory \cite{Gruneisen}.
The method is to use Eq.(\ref{eq:Fqp}) for the free energy, calculate $P(V,T)=-\partial F/\partial V$, and then
find the volume $V(T)$ at which the pressure is zero. 
It is convenient to have a notation for the dimensionless volume expansion, $\zeta$,
\begin{equation}
\zeta=(V-V_0)/V_0
\label{eq:zeta}
\end{equation}
where $V_0$ is as usual the volume that minimizes the frozen lattice (Born-Oppenheimer) energy.
For results to order $\epsilon^2$, it is necessary to know the DFT frozen lattice energy $E_{\rm el}$ to third
order in $\zeta$,
\begin{equation}
E_{\rm el}(V)=E_{\rm el}(V_0) + \frac{1}{2} B_0 V_0\zeta^2 + \frac{1}{6}\dot{B}_0 V_0\zeta^3 +\cdots
\label{eq:Eel}
\end{equation}
Here $B_0 = V_0 (d^2 E_{\rm el}/dV^2)_{V_0}$ is the order $\epsilon^0$ electronic contribution to the bulk modulus,
and ${\dot B}_0$ is the third derivative, $V_0^2 (d^3 E_{\rm el}/dV^3)_{V_0}$.   $\dot{B}_0$ is similar in order of magnitude
to $B_0$.  The notation $\dot{B}$ is used
because the notation $B^\prime$ means $dB/dP=-1-\dot{B}/B$ in equation-of-state theory.\cite{OAnderson,DAnderson}  
Normally $\dot{B}_0<0$ is found.  Crystals are softer when expanded and stiffer when compressed.  
From the volume derivative, we get an ``equation of state,''
\begin{eqnarray}
P(V,T)&=&-B_0\zeta - \frac{1}{2} \dot{B}_0 \zeta^2 
-\sum_Q\frac{\partial F_H}{\partial \omega_Q} \frac{\partial \omega_Q}{\partial V} \nonumber \\
&+&\frac{\hbar}{2}\sum_Q \left[ \frac{\partial \Delta_Q^{(2)}}{\partial V}(n_Q+1/2)
+\Delta_Q^{(2)} \frac{\partial n_Q}{\partial\omega_Q} \frac{\partial\omega_Q}{\partial V} \right].\nonumber \\
\label{eq:PVT}
\end{eqnarray}
Making the substitutions $\partial F_H/\partial\omega_Q = \hbar(n_Q +1/2)$ and
$\partial n_Q/\partial \omega_Q =-(\hbar/k_B T)n_Q (n_Q+1)$, and setting $P=0$, this becomes
\begin{eqnarray}
\zeta&=& -\frac{\dot{B}_0}{2B_0}\zeta^2 +\sum_Q \frac{\hbar\omega_Q}{B_0 V} (\gamma_Q - \delta_Q /2)(n_Q+1/2)
 \nonumber \\
&+&\frac{1}{2}\sum_Q \frac{\hbar\omega_Q}{B_0 V} \frac{\hbar\Delta_Q^{(2)}}{k_B T} \gamma_Q n_Q (n_Q +1).
\label{eq:VV}
\end{eqnarray}
Here the definitions have been introduced
\begin{equation}
\gamma_Q=-\frac{V}{\omega_Q} \frac{\partial \omega_Q}{\partial V},
\label{eq:grun}
\end{equation}
\begin{equation}
\delta_Q=-\frac{V}{\omega_Q} \frac{\partial \Delta_Q^{(2)}}{\partial V},
\label{eq:delta}
\end{equation}
where $\gamma_Q$ is the ``mode Gr\"uneisen parameter,'' and $\delta_Q$ is the analogous volume derivative
of the correction $\omega_Q -\omega_{QH}$.  Therefore, $\delta_Q=\gamma_Q-(\omega_{QH}/\omega_Q)\gamma_{QH}$,
where 
\begin{equation}
\gamma_{QH}=-\frac{V}{\omega_{QH}} \frac{\partial \omega_{QH}}{\partial V}.
\label{eq:gammaH}
\end{equation}

\subsection{Lowest-order (Gr\"uneisen) theory}
\label{subsec:a}

Gr\"uneisen parameters $\gamma_Q$ are of order 1, while $\Delta_Q^{(2)}$ is a small anharmonic correction
of order $\epsilon \omega_Q$.  Therefore $\delta_Q$ is of order $\epsilon\gamma_Q$. 
All the terms on the right of Eq.(\ref{eq:VV}) except the first term involving $\gamma_Q$ are higher order 
in $\epsilon$.  Therefore, the leading-order relation for the 
thermal shift of the volume ($\zeta\equiv(V-V_0)/V_0$) is
\begin{equation}
\zeta_G\equiv\left(\frac{V-V_0}{V_0}\right)_G=\sum_Q \left(\frac{\hbar\omega_{Q}}{B_0 V}\right) 
\gamma_{Q} \left(n_{Q}+\frac{1}{2}\right),
\label{eq:deltaV}
\end{equation}
\begin{equation}
\alpha_G \equiv\frac{1}{V}\left( \frac{\partial V}{\partial T} \right)_G= \frac{1}{B_0}\sum_Q C_{QH}(T)\gamma_Q,
\label{eq:alphaG}
\end{equation}
where $C_{QH}(T)=(\hbar\omega_Q/V)\partial n_Q/\partial T$ is the specific heat per harmonic mode.
These are the famous Gr\"uneisen relations.  Gr\"uneisen's papers \cite{Gruneisen} of 1912 and 1918 were a remarkable advance, simultaneous with the first true
understanding of crystals that came from Rutherford and Bohr, von Laue and the Braggs, Born and
von Karman, Eucken, and Debye.  
Geophysicists and others like to define an average Gr\"uneisen parameter $\gamma$
and to write Eq.(\ref{eq:PVT}) as $P=P_{\rm el}+\gamma U_{\rm vib}/V$ where $U_{\rm vib}$ 
is the harmonic vibrational energy, Eq.(\ref{eq:UH}).  This is called the ``Gr\"uneisen equation of state."   
It omits the higher-order corrections which are now to be discussed.

\subsection{Quasi-harmonic theory}
\label{subsec:QH}

The need for corrections is evident from Fig. \ref{fig:alpha}, showing large high-$T$ deviations
of thermal expansion relative to specific heat.  If Eq.(\ref{eq:alphaG}) were correct, this would
require unphysically large $T$-dependence of Gr\"uneisen constants.
To correctly find the volume shift $\zeta$ to next order, it is necessary to solve Eq.(\ref{eq:VV}) self-consistently.
The job is complicated by the fact that $\gamma_Q$ as defined in Eq.(\ref{eq:grun}), and 
$\gamma_{QH}$ as defined in Eq.(\ref{eq:gammaH}), are volume dependent.  

An interesting example is the computation by Skelton {\it et al.} \cite{Skelton} of $\alpha(T)$ for PbS, PbSe,
and PbTe.  Above the Debye temperature, $\alpha$ shows strong $T$-dependence, similar to argon in Fig. \ref{fig:alpha}.
These computations had no anharmonic corrections.  This shows that a good quasi-harmonic theory
does in fact have important corrections beyond the lowest order Gr\"uneisen theory.  The volume-dependent
electron and vibrational free energy in Eq.(\ref{eq:Fqp}) are included, omitting the correction term involving 
$\Delta_Q^{(2)}$.  The resulting $F_{QH}(V,T)$ is minimized at fixed $T$, giving $V_{QH}(T)$.  This is equivalent
to a self-consistent solution of Eq.(\ref{eq:VV}), omitting the terms involving $\delta_Q$ and $\Delta_Q^{(2)}$. 

\subsection{Full second-order theory}
\label{subsec:c}

It is rather messy to do the full solution to second order.  To simplify things, consider just the high-$T$ limit, where
$n_Q + 1/2$ is $\approx k_B T/\hbar\omega_Q$ and $n_Q(n_Q + 1)$ is $\approx(k_B T/\hbar\omega_Q)^2$.  
The quantum corrections are factors $[1\pm(\hbar\omega/k_B T)^2/12 + 4^{\rm th} {\rm order}+\cdots]$.
Then, to order $\epsilon^2$, Eq.(\ref{eq:VV}) becomes
\begin{eqnarray}
\zeta&=& -\frac{\dot{B}_0}{2B_0}\zeta_0^2 \nonumber \\
&+& \frac{k_B T}{B_0 V_0 (1+\zeta_0)}  \sum_Q \left( \gamma_Q(V) - \frac{\delta_{Q0}}{2}
+\frac{\Delta_{Q0}^{(2)}}{2\omega_{Q0}} \gamma_{Q0} \right).
\label{eq:VVT}
\end{eqnarray}
The subscript 0's indicate that quantities are all (except the first appearance of $\gamma_Q$) evaluated at the
frozen-lattice $T=0$ minimum volume $V_0$.  
The lowest-order volume expansion, $\zeta_0$, at high $T$, from Eq.(\ref{eq:deltaV}) or from Eq.(\ref{eq:VVT}) is
\begin{equation}
\zeta_0=\frac{k_B T}{V_0 B_0} \sum_Q\gamma_{Q0}.
\label{eq:zeta0}
\end{equation}
The definition of $\gamma_{Q0}$ is
\begin{equation}
\gamma_{Q0}=-\frac{V_0}{\omega_{Q0}}\left( \frac{\partial\omega_{QH}}{\partial V}\right)_{V_0},
\label{eq:gamma0}
\end{equation}
and the phonon frequencies $\omega_{Q0}$ are similarly the $T=0$ harmonic frozen lattice values. 

We now need to expand the fully anharmonic $\gamma_{Q}$ around $\gamma_{Q0}$, to first order in $\zeta$.
This is done in two stages.  First we expand around $\gamma_{QH}$,
\begin{eqnarray}
\gamma_Q(V)&\equiv& -\frac{V}{\omega_{QH}+\Delta_Q^{(2)}} \frac{\partial(\omega_{QH}+\Delta_Q^{(2)})}{\partial V} 
\nonumber \\
&\approx& \left( 1-\frac{\Delta_{Q0}^{(2)}}{\omega_{Q0}} \right) \gamma_{QH}(V) + \delta_{Q0}
\label{eq:g1}
\end{eqnarray}
Subscripts 0 indicate sufficient accuracy for a first-order result.
Next, we expand $\gamma_{QH}(V)$ around the volume $V_0$ where $\gamma_{Q0}$ is defined.
To do this, we need to know the harmonic frequencies to second order in $\zeta$ around $V_0$,
\begin{equation}
\omega_{QH}(V)=\omega_{Q0}\left[ 1 - \gamma_{Q0}\zeta -\dot{\gamma}_{Q0}\zeta^2/2+\cdots\right],
\label{eq:gamdot}
\end{equation}
\begin{equation}
\dot{\gamma}_{Q0}\equiv-\frac{V_0^2}{\omega_{Q0}}\left( \frac{\partial^2\omega_{QH}}{\partial V^2} \right)_{V_0}.
\label{eq:dotgrun}
\end{equation}
The notation $\dot{\gamma}_Q$ 
used here (Eq.\ref{eq:dotgrun}) is not the same as
$V(\partial \gamma_Q/\partial V) = \gamma_Q + \gamma_Q^2 + \dot{\gamma}_Q$.
Volume dependence of $\gamma_Q(V)$ has often been neglected.
If the mode Gr\"uneisen parameter were independent of volume, one could integrate
to find $\omega_Q(V)=\omega_Q(V_0)(V/V_0)^{\gamma_Q}$.  
As observed previously \cite{Nielsen,Kunc}, there is no justification for this except unwarranted
optimism.   After some algebra, the relation between $\gamma_{QH}$ and $\gamma_{Q0}$ is
\begin{equation}
\gamma_{QH}(V)=\gamma_{Q0}+\left[ \gamma_{Q0}+\gamma_{Q0}^2+\dot{\gamma}_{Q0} \right] \zeta.
\label{eq:gamH}
\end{equation}
Combining the two stages, the result is
\begin{equation}
\gamma_Q = \gamma_{Q0} + \left[ \gamma_{Q0}+\gamma_{Q0}^2+\dot{\gamma}_{Q0} \right] \zeta_0
-\frac{\Delta_Q^{(2)}}{\omega_{Q0}}\gamma_{Q0} +\delta_{Q0} + \cdots.
\label{eq:gamQ}
\end{equation}
Finally, insert this into Eq.(\ref{eq:VVT}) and keep only first order corrections.  The result is
\begin{eqnarray}
\zeta &=& \zeta_0 -\frac{\dot{B}_0}{2B_0} \zeta_0^2 \nonumber \\
&+& \frac{k_B T}{B_0 V_0}  \sum_Q \left[\left( \gamma_{Q0}^2+\dot{\gamma}_{Q0}\right)\zeta_0
+\frac{\delta_{Q0}}{2} -\frac{\Delta_{Q0}^{(2)}}{2\omega_{Q0}}\gamma_{Q0}  \right] 
 \nonumber \\
\label{eq:Vfinal}
\end{eqnarray}
The corresponding high $T$ formula for the volume thermal expansion coefficient is
\begin{equation}
\alpha = \alpha_0 +\Delta\alpha_{QH} + \Delta\alpha_{AH}.
\label{eq:alphatot}
\end{equation}
The leading term, $\alpha_0$, is just the Gr\"uneisen formula evaluated with frozen-lattice parameters.
Its high-$T$ form is
\begin{equation}
\alpha_0=\frac{k_B}{B_0 V_0} \sum_Q \gamma_{Q0}.
\label{eq:alpha0}
\end{equation}
The high-$T$ quasiharmonic correction is explicitly linear in $T$,
\begin{equation}
\Delta\alpha_{QH}= 2T\alpha_0 \left( \frac{k_B}{B_0 V_0} \sum_Q
(\gamma_{Q0}^2 + \dot{\gamma}_{Q0})\right) -\frac{\dot{B}_0 T}{B_0} \alpha_0^2.
\label{eq:alphaQH}
\end{equation}
These are smaller than the leading term $\alpha_0$ by a factor like $T\alpha_0 \sim \epsilon$.
The high-$T$ anharmonic correction is
\begin{equation}
\Delta\alpha_{AH}=\frac{k_B}{B_0 V_0}\sum_Q \left(\delta_{Q0} 
-\frac{\Delta_{Q0}^{(2)}}{\omega_{Q0}}\gamma_{Q0} \right).
\label{eq:alphaAH}
\end{equation}
This is also smaller than $\alpha_0$ by one power of $\epsilon$.
The anharmonic factors $\delta-(\Delta/\omega)\gamma$ vary linearly with $T$ at high $T$.  This is why the factors of 1/2 
multiplying $\delta_{Q0}$ and $(\Delta_{Q0}^{(2)}/\omega_{Q0})\gamma_{Q0}$ 
in Eq.(\ref{eq:Vfinal}) disappear in Eq.(\ref{eq:alphaAH}) after taking
the temperature derivative.

The QH calculations of Karki {\it et al.} \cite{Karki} for MgO show that QH corrections, Eq.(\ref{eq:alphaQH}),
can cause  a large effect, even exceeding 
the experimental linear rise in $\alpha$.
The calculations of Mounet and Marzari \cite{Mounet} also show a significant QH
linear increase of $\alpha(T)$ in diamond, but less than the shift observed experimentally by Slack and Bartram \cite{Slack}.
These results indicate that the anharmonic part of Eq.(\ref{eq:alphatot}) is as important as the QH part.
A path integral Monte Carlo study by Herrero and Ramirez \cite{Herrero} confirms this.

\section{Thermal correction to Bulk Modulus} \label{sec:BM}

The literature about $B(T)$ is large because of its importance in geoscience.
The bulk modulus is
 the simplest and most accessible part of the elastic constant matrix $C_{ij}$, all components of which show 
related zero-point and thermal alteration.  This paper focuses on $B$ for simplicity, but generalization to the
 full elasticity tensor is not hard. \cite{Leibfried,WallaceOld,Wang,Mounet,Karki,Paszkiewicz,Schiferl,
 Karch,Greeff,Liu,Shao,Steneteg}  The bulk modulus is dominated by the large electronic
 contribution $B_0$.  Corrections to first and second order in $\epsilon$ can be found from the
 quasiparticle free energy, Eq.(\ref{eq:Fqp}), and pressure, Eq.(\ref{eq:PVT}).
The first-order shift of the isothermal bulk modulus uses only the first two correction terms in Eq.(\ref{eq:PVT}),
\begin{equation}
P(V,T)=-B_0\zeta - \frac{1}{2} \dot{B}_0 \zeta^2 
-\sum_Q\frac{\hbar\omega_Q}{V} \gamma_Q (n_Q+1/2)
\label{eq:P1}
\end{equation}
There is no anharmonic contribution to $B=-V\partial P/\partial V$ in first order.
The simulations by Ramirez {\it et al.} \cite{Ramirez} (using a fluctuation formula\cite{Landau0} for $B(T)$ ) confirm 
the accuracy of quasiharmonic theory for phases of ice. 
For cases like ice, where volume shifts are relatively large, it is insufficient to compute only low-order
derivatives of energy and vibrational frequency.    But QH theory, with $\gamma_Q(V)$ computed separately for
different volumes along the QH $V(T)$-curve, has been shown to work. \cite{Ramirez} 

Taking the volume derivative of Eq.(\ref{eq:P1}),
\begin{eqnarray}
&&B_{T,QH}(T)=\frac{V}{V_0}(B_0+\dot{B}_0 \zeta) \nonumber \\
&&- B_0\sum_Q \left(\frac{\hbar\omega_Q}{B_0 V}\right) \left[
T \left( \frac{\partial n_Q}{\partial T} \right)_{\rm H} \gamma_Q^2 
+  \left( n_Q + \frac{1}{2} \right) \dot{\gamma}_Q \right]. \nonumber \\
\label{eq:Beos}
\end{eqnarray}
The last term uses the
identity $\partial(\omega_Q \gamma_Q /V)/\partial V = \omega_Q \dot{\gamma}_Q/V^2$.
The next to last term uses the fact that $V\partial n_Q/\partial V$ equals $T(\partial n_Q/\partial T)_{\rm H} \gamma_Q$.
The first term of Eq.(\ref{eq:Beos}) is the purely electronic term, $B_{\rm el}$.
To first order in $\epsilon$ it can be written as $B_0 + (B_0 + {\dot B}_0)\zeta_0$.
The high $T$ version of $\zeta_0$ is given in Eq.(\ref{eq:zeta0}), and the general expression
is the same as the Gr\"uneisen version, Eq.(\ref{eq:deltaV}), except frequencies and derivatives are
evaluated at $(V,T)=(V_0,0)$.  Then Eq.(\ref{eq:Beos}) becomes
\begin{eqnarray}
\left(\frac{\Delta B}{B_0}\right)_T &=&
\sum_Q  \left(\frac{\hbar\omega_{Q0}}{B_0 V_0} \right)
\left[ -T\left(\frac{\partial n_{Q0}}{\partial T}\right)_{\rm H}\gamma_{Q0}^2 \right. \nonumber \\ 
&+& \left. \left(n_{Q0}+\frac{1}{2}\right) \left(\gamma_{Q0}\left[1+\frac{\dot{B}_0}{B_0}\right]
-\dot{\gamma}_{Q0}\right)\right],
\label{eq:Bfinal}
\end{eqnarray}
This equation is contained in somewhat hidden form in Leibfried and Ludwig \cite{Leibfried}.  
Born and Huang \cite{Born} and Wallace \cite{WallaceOld}
also give this result, except altered because frequencies and derivatives are evaluated at $(V,T)$.
The paper of Karch {\it et al.},\cite{Karch} gives an alternate derivation.  Many incorrectly simplified 
versions exist.  

The parameter $\epsilon$ is not truly small for ice Ih.  For this reason,
Eq. \ref{eq:Bfinal} does not work particularly well \cite{Pamuk,Ramirez}.  Direct computation and 
minimization of the QH free energy (Eq. \ref{eq:Fqp} without the last term) may work.
This has done used for many years, even in cases where the shifts are small enough that Eq.(\ref{eq:Bfinal}) should be adequate.\cite{Cowley,Schiferl,Greeff,Mounet,Wang,Dewaele}   In cases, like ice Ih, where
$\epsilon$ is too large to use Eq.(\ref{eq:Bfinal}), there is no guarantee that truly anharmonic terms of order $\epsilon^2$ and
higher are not as important as QH terms found by direct minimization.

It is important to distinguish between adiabatic ($B_S$) and
isothermal ($B_T$) conditions \cite{Paszkiewicz}.  The definitions are
\begin{equation}
B_T = -V(\partial P/\partial V)_T = V(\partial^2 F/\partial V^2)_T
\label{eq:BT}
\end{equation}
\begin{equation}
B_S = -V(\partial P/\partial V)_S = V(\partial^2 U/\partial V^2)_S
\label{eq:BS}
\end{equation}
where $U$ and $F$ are the internal energy and Helmholz free energy respectively.
Thermodynamics gives exact relations \cite{Leibfried,OAnderson,Wang,Paszkiewicz,DAnderson,WallaceOld,Davies,Landau,Wehner},
\begin{equation}
\frac{B_S }{ B_T} -1 =\frac {C_P}{C_V} -1 = \frac{ T \alpha^2 B_T V}{C_V} =\frac{T \alpha^2 B_S V}{C_P},
\label{eq:reln}
\end{equation}
where $C_V/V$ is the heat capacity per volume.  The product $\alpha T$ is of order $\epsilon$,
and $C_V T/B_T V$ is also of order $\epsilon$, so the shift $(B_S-B_T)/B$ is positive and order $\epsilon$. 
The full tensor version also is available \cite{oldWallace,WallaceOld,Wehner,Hearmon}.
The vibrational corrections $\delta_S$ (adiabatic) and $\delta_T$ (isothermal) are both first order in $\epsilon$, and
they differ from each other in the same order.  The leading order value of $T \alpha^2 B_T V/C_V$ is sufficient for 
correcting isothermal to adiabatic.    Using Eq.(\ref{eq:alphaG}) and the harmonic specific heat, the result is
\begin{equation}
B_S - B_T = \frac{T}{V} \frac{\left[\sum_Q \hbar\omega_{Q0} \left(\frac{\partial n_{Q0}}{\partial T}\right)_{\rm H}
\gamma_{Q0}\right]^2}{\sum_Q  \hbar\omega_{Q0} \left(\frac{\partial n_{Q0}}{\partial T}\right)_{\rm H}}.
\label{eq:BTtoBS}
\end{equation}
Figure \ref{fig:data} shows approximate high $T$ slopes ($dB/dT$) of both $B_S$ and $B_T$ for NaCl.  
In the high $T$ limit where $\hbar\omega_Q(dn_Q/dT)\rightarrow k_B$, Eq.(\ref{eq:BTtoBS}) reduces
to $d(B_S - B_T)/dT = 3k_B \bar{\gamma}^2 /V_a$, where $\bar{\gamma}=\sum_Q \gamma_Q/3N$.
The slopes shown in Fig. 1 then require $\bar{\gamma}\approx 1.5$,
in good agreement with other estimates of $\bar{\gamma}$ for NaCl.

\begin{appendix} 
\section{}
\label{app:A}

This appendix tries to illuminate the quasiparticle thermodynamics of
Eqs.(\ref{eq:Shat}-\ref{eq:anhsh}) by using anharmonic
thermal perturbation theory.
According to Cowley \cite{oldCowley}, the vibrational thermal Green's function matrix, in the basis of 
harmonic eigenstates $|\lambda\rangle$ with frequencies $\omega_{\lambda,H}$, is
\begin{equation}
(G^{-1})_{\lambda\lambda^\prime} =
(\omega_{\lambda,H}^2 -\omega^2) \delta_{\lambda\lambda^\prime}+2(\omega_\lambda \omega_{\lambda^\prime})^{1/2}
[\Delta_{\lambda\lambda^\prime}-i\Gamma_{\lambda\lambda^\prime}]
\label{eq:G1}
\end{equation}
The eigenvalues $\omega^2$ of the matrix $\hat{G}^{-1}+\omega^2 \hat{1}$, with the imaginary part $\hat{\Gamma}$ omitted,
are denoted  $\omega_\lambda^2$.  They are the renormalized squared normal mode frequencies.  If anharmonicity is weak,
then in leading approximation these eigenvalues are $\omega_\lambda^2 = \omega_{\lambda,H}^2
+2\omega_\lambda \Delta_{\lambda\lambda}$.  At the same level of approximation, 
$\omega_\lambda=\omega_{\lambda,H}+\Delta_{\lambda\lambda}$.  Cowley gives an explicit formula
from lowest-order perturbation theory, for the anharmonic shift $\Delta_{\lambda\lambda}^{(2)}=
\Delta_Q^{(2)} =\omega_Q - \omega_{Q,H}$.
The normal mode index $\lambda$ is now replaced by $Q=(\vec{Q},j)$.  Cowley's formula can be written
in the form
\begin{equation}
\Delta_Q^{(2)} = \frac{1}{N}\sum_{Q^\prime} \frac{\partial \omega_Q}{\partial n_{Q^\prime} } (n_{Q^\prime}+1/2) ,
\label{eq:G2}
\end{equation}
\begin{eqnarray}
\frac{\partial \omega_Q}{\partial n_{Q^\prime} } &=&\frac{24}{\hbar} V^{(4)} (QQ,Q^\prime Q^\prime)\nonumber \\
&-&\frac{36}{\hbar^2}\sum_{Q^{\prime\prime}} |V^{(3)}(Q Q^\prime Q^{\prime\prime})|^2 \left[
\frac{1}{\omega_{Q^{\prime\prime}}+\omega_{Q^\prime}+\omega_Q} +  \right. \nonumber \\
&&\frac{1}{\omega_{Q^{\prime\prime}}+\omega_{Q^\prime}-\omega_Q}  +
\frac{1}{\omega_{Q^{\prime\prime}}-\omega_{Q^\prime}+\omega_Q} + \nonumber \\
&& \left. \frac{1}{\omega_{Q^{\prime\prime}}-\omega_{Q^\prime}-\omega_Q}  \right].
\label{eq:G3}
\end{eqnarray}
This is an explicit form for Eq.(\ref{eq:anhsh}).
Here $V^{(3)}$ and $V^{(4)}$ are third and fourth derivatives of the BO potential taken around the
periodic sites of the crystal of volume $V_0$, and the frequencies $\omega_Q$ and the occupation
number $n_{Q^\prime}$ use anharmonic renormalization (computed self-consistently.)

Cowley also derives the anharmonic free energy at the same level of perturbation theory.  His answer
can be written as
\begin{equation}
F = F_{H,0} + \frac{\hbar}{2}\sum_Q \Delta_Q^{(2)} (n_Q + 1/2) + F_{A0} 
\label{eq:G4}
\end{equation}
\begin{eqnarray}
F_{A0}&=&-\frac{3}{2\hbar} \sum_{Q Q^\prime Q^{\prime\prime}} 
|V^{(3)}(Q Q^\prime Q^{\prime\prime})|^2 \times \nonumber \\
&&\left[ \frac{1}{\omega_{Q^{\prime\prime}}+\omega_{Q^\prime}+\omega_Q} -
\frac{3}{\omega_{Q^{\prime\prime}}+\omega_{Q^\prime}-\omega_Q} \right]
\label{eq:G5}
\end{eqnarray}
where $F_{H,0}$ is the free energy of non-interacting (harmonic) quasiparticles.
Now find the corresponding entropy, $S=-dF/dT$.  The non-interacting part gives the
harmonic entropy,
\begin{equation}
S_{H,0}=k_B \sum_Q [(n_Q +1)\ln(n_Q +1) - n_Q \ln n_Q ]
\label{eq:G6}
\end{equation}
Consider what happens if the ``quasiparticle entropy" is constructed by replacing the
harmonic frequencies in Eq.(\ref{eq:G6})  by
the anharmonic frequencies $\omega_{Q,H}+\Delta_Q^{(2)}$. Taylor expanding
to first order in $\Delta_Q$, the answer is
\begin{equation}
\Delta S = S_{QP}-S_{H,0} = -\hbar\sum_Q \frac{\partial n_Q}{\partial T} \Delta_Q^{(2)}.
\label{eq:G7}
\end{equation}
This is the same as the entropy $dF/dT$ from Eq.(\ref{eq:G4}).  The factor $1/2$ in \ref{eq:G4} disappears
when using Eq.(\ref{eq:G2}) while differentiating Eq.(\ref{eq:G4}) for $\Delta F$, because
$\partial \omega_Q / \partial n_{Q^\prime} $ is symmetric in $Q$ and $Q^\prime$.
An alternate derivation using
a variational principle is given in ref. \onlinecite{Hui}.  This suggests that the use of QP energies
in the harmonic entropy formula may be  valid somewhat 
beyond low-order perturbation theory.

Consider then what happens if the same substitution $\omega_{Q,H} \rightarrow \omega_Q$
is done in the harmonic free energy $F_H = \sum_Q \hbar\omega_{Q,H}(n_Q + 1/2) - TS_H$.  
The answer is $\Delta F_H = \hbar\sum_Q \Delta_Q^{(2)} (n_Q + 1/2) -T\Delta S.$  This differs from
the correct anharmonic free energy, Eq.(\ref{eq:G4}), by not having the correct factor of $1/2$.  This is a
proof of the double-counting correction that was added to the internal energy in Eq.(\ref{eq:UQP}).
The correct formula Eq.(\ref{eq:G4}) does differ from the QP theory of Eqs.(\ref{eq:Shat}-\ref{eq:anhsh}) 
by a small $T$-independent term $F_{A0}$.

\end{appendix}

\acknowledgements

I thank B. Pamuk and M. Fern\'andez-Serra, whose calculations for ice inspired this paper.
I thank A. G. Abanov for discussions that lead to Eq.(\ref{eq:UQP}).
I thank R. Lieberman and D. Weidner for help with the geophysical literature, and V. Pokrovsky 
for help with the physics literature.  I thank D. A. Broido, H. Huang,
C. Marianetti, and R. Wentzcovitch for useful conversations, and C. Herrero, and R. Ramirez for helpful correspondence.
I especially thank A. Mayer for a very helpful correspondence and a careful reading of the manuscript.
This work was supported in part by DOE grant No. DE-FG02-08ER46550.

\bibliography{citation}
\bibliographystyle{unsrt}

\end{document}